\documentclass[trans]{IEEEtran}
\usepackage{cite}
\usepackage{amsmath,amssymb,amsfonts}
\usepackage{algorithmic}
\usepackage{textcomp}
\usepackage{multicol}
\usepackage{epsfig}
\usepackage{lipsum}
\usepackage{booktabs,tabularx}
\usepackage{longtable}
\usepackage{vcell}
\usepackage{wrapfig}
\usepackage{algorithm}
\usepackage{epsfig}
\usepackage{graphicx}
\usepackage{booktabs}
\usepackage{multirow}
\usepackage{hhline}
\usepackage{textcomp}
\usepackage{url}
\usepackage[leftcaption]{sidecap}
\usepackage{caption}
\usepackage{xcolor, soul}
\usepackage{colortbl}
\usepackage{caption}
\usepackage{subfig}
\usepackage{booktabs}
\usepackage{multirow}
\usepackage{graphicx}
\usepackage{epsfig}

\begin{document}

\title{Communication and Control in Collaborative UAVs:  Recent Advances and Future Trends}
\author{Shumaila Javaid, Nasir Saeed, ~\IEEEmembership{Senior Member,~IEEE,} Zakria Qadir, ~\IEEEmembership{Student Member,~IEEE,} Hamza Fahim, Bin He, Houbing
Song, ~\IEEEmembership{Senior Member,~IEEE,} and Muhammad Bilal, ~\IEEEmembership{Senior Member,~IEEE}
\thanks{S. Javaid, H. Fahim, and B. He are with the Department of Control Science and Engineering, College of Electronics and 	Information Engineering, Tongji University, Shanghai 201804, China, and also with Frontiers 	Science Center for Intelligent Autonomous Systems, Shanghai 201210. Email: \{shumaila, hamzafahim, hebin\}@tongji.edu.cn.\newline
N. Saeed is with the Department of Electrical and Communication Engineering, United Arab Emirates University (UAEU), Al Ain, 15551, UAE. Email: mr.nasir.saeed@ieee.org.\newline
Z. Qadir is with the School of Engineering, Design and Built Environment, Western Sydney University, Locked Bag 1797, Penrith, NSW 2751, Australia. Email: z.qadir@westernsydney.edu.au. \newline
H. Song is with Embry-Riddle Aeronautical University, Daytona Beach, FL
32114 USA. Email: h.song@ieee.org.\newline
M. Bilal is with the Department of Computer Engineering, Hankuk University
of Foreign Studies, Yongin-si, Gyeonggido, 17035, South Korea. Email:
m.bilal@ieee.org.}}

\maketitle
\begin{abstract}
The recent progress in unmanned aerial vehicles (UAV) technology has significantly advanced UAV-based applications for military, civil, and commercial domains. Nevertheless, the challenges of establishing high-speed communication links, flexible control strategies, and developing efficient collaborative decision-making algorithms for a swarm of UAVs limit their autonomy, robustness, and reliability. Thus, a growing focus has been witnessed on collaborative communication to allow a swarm of UAVs to coordinate and communicate autonomously for the cooperative completion of tasks in a short time with improved efficiency and reliability. This work presents a comprehensive review of collaborative communication in a multi-UAV system. We thoroughly discuss the characteristics of intelligent UAVs and their communication and control requirements for autonomous collaboration and coordination. Moreover, we review various UAV collaboration tasks, summarize the applications of UAV swarm networks for dense urban environments and present the use case scenarios to highlight the current developments of UAV-based applications in various domains. Finally, we identify several exciting future research direction that needs attention for advancing the research in collaborative UAVs.
\end{abstract}
\IEEEpeerreviewmaketitle


\section{Introduction}
\label{sec:introduction}
Unmanned Ariel Vehicles (UAVs) ability to fly at a controlled speed and height to carry out designated tasks gained significant recognition in military operations for long-term surveillance, decoys, and missile launches against fixed targets \cite{zhang2019survey, ullah2020cognition,sharma2020communication,haider2022novel}. Nevertheless, recently the tremendous potential of UAVs also surged their exploration for various civilian applications, including transportation, recuse, disaster relief, wireless recovery, and agriculture \cite{torresan2017forestry,shakhatreh2019unmanned,pajares2015overview}. Most of these UAV-based applications require a swarm of UAVs rather than a single UAV to work together to perform independent operations to complete tasks. However, in a multi-UAV system, the lack of proper UAV-to-UAV communication architecture influences their performance in a distributed environment. The multi-UAV communication network architecture allows UAVs to perform different tasks collaboratively by sharing information and responsibilities. Furthermore, collaborative UAV communication enables fast operations with more reliability and robustness; for instance, if a specific UAV fails to perform some task due to failure, other UAVs can continue to complete the job \cite{erdelj2017wireless,shakeri2019design}. Therefore, inspired by the great prospective of collaborative UAV communication, scientific and research communities focused on investigating the possibilities and limitations of achieving collaborative UAV communication \cite{chmaj2015distributed,sun2017collision,dinh2019joint}.

State-of-the-art literature has presented various collaborative communication schemes for UAV swarm networks in different domains to optimize service time, energy efficiency, coverage and communication performance. For example, in the context of UAV-based collaborative beamforming \cite{ sun2022collaborative},  UAVs can realize collaborative beamforming for establishing a virtual antenna array to generate a beam pattern with a sharp main lobe and low sidelobe levels to enhance the antenna gain, reduce interference and improve the signal-to-noise ratio of the received signal and focused transmitted signal \cite{sun2021time}. Similarly, in collaborative multiple-target tracking \cite{zhou2021intelligent}, the natural characteristics of flexible mobility of UAVs can play a significant role in sensing and tracking mobile targets at a large scale, leading to advanced disaster monitoring, damage assessment, manufacturing safety, and border security \cite{ xiao2019coverage, zhang2019iot}. At the same time, collaborative UAV routing also offers efficient ways for tasks offloading in a distributed manner, such as localization, actuation-based task assignments and optimal path selection from source to destination for different product delivery and disaster relief applications \cite{mukherjee2020ecor, khan2019routing}. However, a few primary constraints include lack of physical infrastructure, high mobility, channel characterization, intermittent connectivity, bounded transmission range, and limited resources \cite{jawhar2017communication,chmaj2015distributed} are hindering the development of collaborative UAV swarm architecture.

Accordingly, existing studies address the aforementioned issues by focusing on different solutions, including channel characterization, resource management, data communication, and emerging technologies integration (such as 5G and 6G) to enable fast and reliable collaborative UAV communications. For example, in \cite{ li2018uav}, Li et al. highlighted the advancement of 5G-assisted UAV communication for achieving high reliability, fast speed, rapid recovery, flexibility, and cost-effective traffic offloading in highly crowded areas. In another work \cite{ zeng2019accessing}, authors identified the limitations of UAV communication (such as Line of Sight (LoS) dominant UAV-ground channel, high Quality of Service (QoS) requirements, and inadequate power, size and weight constraints) that can be addressed using 5G and beyond 5G technology. A few recent studies \cite{ zhang2019survey, ullah2020cognition, marchese2019iot,mozaffari2019tutorial,hentati2020comprehensive}, investigated the 5G millimeter-wave communication and the scope of 5G aided UAV communication for channel characterization, standardization, collision avoidance, energy efficiency, and optimal trajectory design.

\begin{figure*}
\center
\includegraphics[width=5in, height = 8cm]{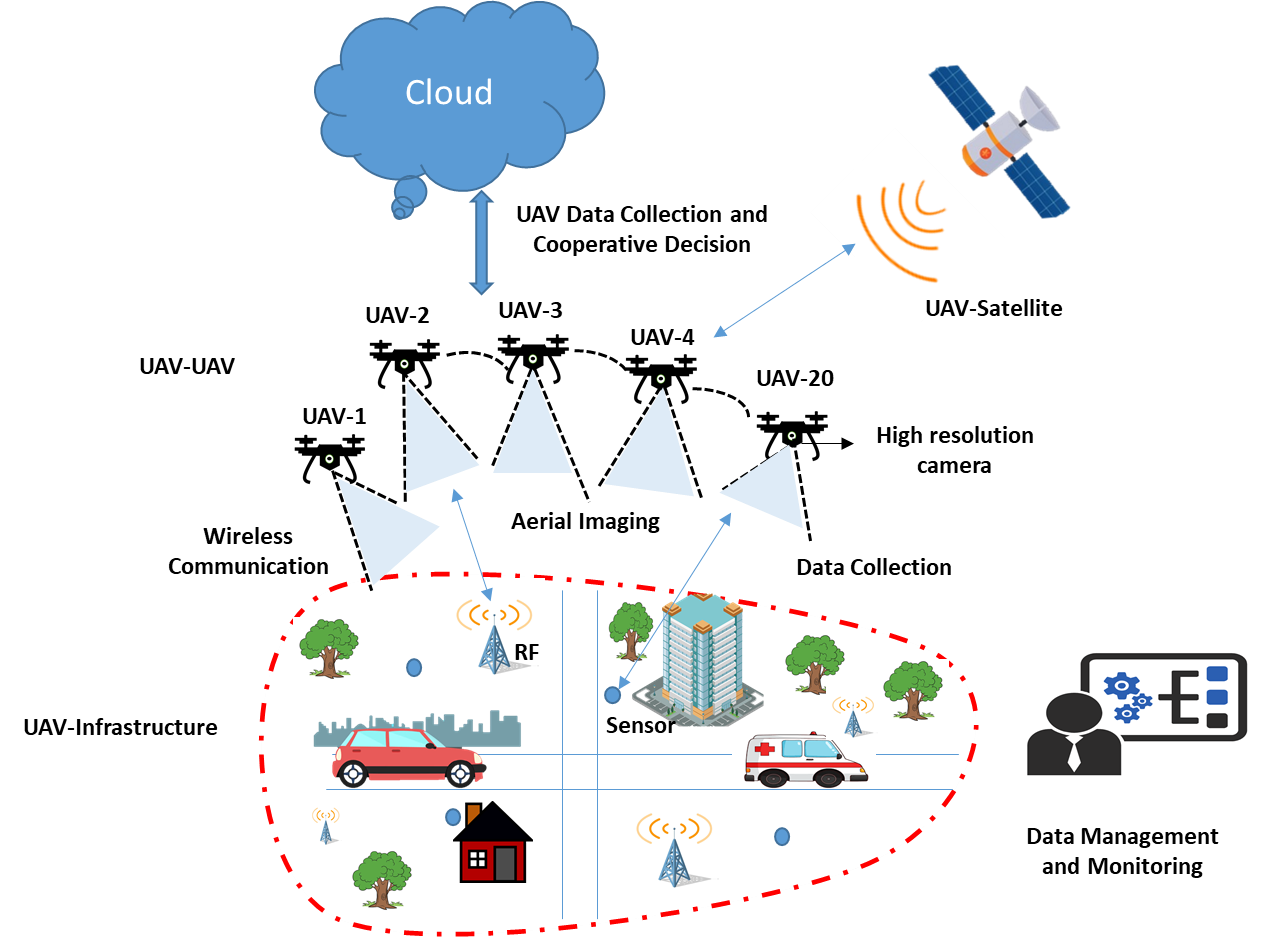}
\vspace{2.5 mm}
\caption{Illustration of a collaborative UAV network.}
\label{Main}
\vspace{-4mm}
\end{figure*}

\subsection{Related Surveys}
Due to emerging applications of UAVs, various surveys in the literature cover different aspects of UAVs. For instance, In  \cite{hentati2020comprehensive}, a detailed review of communication networks and routing protocols are presented that highlight their contribution to improve the reliability, data delivery, and resource optimization of UAV networks. Authors in \cite{ mozaffari2019tutorial} investigated the potential of UAVs for enhancing the performance of wireless networks in terms of coverage, energy efficiency, capacity, and reliability. In another survey \cite{khuwaja2018survey}, Khuwaja et al. reviewed the channel characterization models for UAV communication and discussed various methods that improve the shadowing and propagation effects. In \cite{ yan2019comprehensive} channel fading effects and measurements are thoroughly studied for link budget analysis and reduced link losses. Authors in \cite{ gupta2015survey} highlighted several communication issues such as network setup, link failure, and energy challenges to enable UAV communication. Fotouhi et al. in \cite{fotouhi2019survey} focused on the integration of UAV systems with cellular networks for standardization and advancements of UAV systems.

Furthermore, various state-of-the-art surveys have summarized the primary contributions of existing solutions for achieving collaborative communication. For example, in \cite{ popescu2019survey}, authors review wireless sensor network (WSN) and UAV collaboration for sensor nodes deployment, localization, and UAV's path formation for data collection to enable efficient monitoring in different domains such as agriculture field, environment and disaster area management. The authors explicitly reviewed UAV-WSN architecture for efficient monitoring while lacking focus on UAV-to-UAV collaborative communication for task completion. In another work \cite{chmaj2015distributed}, chmaj et al. presented an overview of different distributed operations such as object tracking, path planning, navigation, monitoring and data manipulation that a swarm of UAVs can perform. However, the authors did not include a comprehensive discussion on collaborative communication architecture and mechanisms for performing distributed operations. Nawaz et al. in \cite{ nawaz2021uav} discussed the characteristics of a UAV network compared to WSN and mobile ad-hoc networks and summarized the network issues (such as routing, power and quality of service) that need to be addressed for collaborative UAV networks. However, the presented review did not discuss the UAV's collaborative tasks in detail. In another work \cite{jawhar2017communication}, authors comprehensively studied the UAV communication links (such as UAV-to-UAV and UAV-to-infrastructure) to identify the requirements of UAV-based communication systems. However, the detailed review of the networking architectures and communication framework did not include the UAV collaboration for joint control and task completion. In \cite{alsamhi2021green}, authors studied the UAV's role in achieving green Internet of Things (IoT) to realize a sustainable smart world. The survey summarizes strategies to integrate UAVs with IoT as edge intelligence devices to collect and process data obtained from IoT devices. The review also explores the opportunities for connectivity and communication beyond 5G. However, it did not include UAV-to-UAV communication for collaborative control and task performance. In another work \cite{ shi2021review }, Shi et al. in reviewed the existing UAV communication protocols for power line inspection industry and classified the UAV communication link type and summarized the wireless mesh networking protocols; however, it lacks focus on the existing schemes related to collaborative UAV communication.

In \cite{ bithas2019survey}, the authors reviewed machine learning techniques for UAV communication and provided an overview for integrating machine learning techniques that can optimize the physical layer and improve resource and network management. Furthermore, Hayat et al. in \cite{ hayat2016survey} studied the civil applications of UAVs and discussed QoS and data communication requirements for multi-UAV communication. While Sharma et al. specifically discussed the cellular communication challenges of UAV-to-UAV and UAV-to-device communication. Table I further highlights the key features of existing related surveys.

Although the surveys mentioned above are quite comprehensive, they lack a comprehensive focus on collaborative communication in UAVs. Therefore, this review aims to fill the gap and provide a detailed study to help design collaborative multi-UAV systems in the future to advance the performance of multi-UAV systems. To this end, this paper discusses the current literature on collaborative UAV communication, including communication and control requirements, use cases for different applications, and opportunities for collaborative UAV communication. Moreover, the paper discusses the open challenges and future research directions that need significant attention to realize the scope of collaborative UAV communication.

% \usepackage{multirow}
% \usepackage{graphicx}
\begin{table}[]
\centering
\label{tab:Relatedsurveys}
\caption{Summary of related surveys}
\begin{tabular}{|p{0.7cm}|p{3.0cm}|p{4.0cm}|}
\hline
\textbf{Ref.} & \textbf{Primary focus} & \textbf{Key features} \\ \hline
\cite{chmaj2015distributed} & \begin{tabular}[c]{@{}l@{}}Distributed operations\\ of UAVs\end{tabular} & \begin{tabular}[c]{@{}l@{}}1. Provides a comprehensive \\ discussion on UAVs swarm \\ network for various operations \\ such as object tracking, path \\ planning, navigation, monitoring\\ and data manipulation\end{tabular} \\ \hline
\cite{jawhar2017communication} & \begin{tabular}[c]{@{}l@{}}UAV communication\\ links\end{tabular} & \begin{tabular}[c]{@{}l@{}}1. Discusses UAV communication \\ links (such as UAV-to-UAV and \\ UAV-to-infrastructure)\end{tabular} \\
 &  & \begin{tabular}[c]{@{}l@{}}2. Study and identify the requirem-\\ ents of UAV-based communication \\ systems\end{tabular} \\ \hline
\multirow{3}{*}{\cite{mozaffari2019tutorial}} & \multirow{3}{*}{\begin{tabular}[c]{@{}l@{}}Prospects of UAV-enabled \\ wireless networks\end{tabular}} & \begin{tabular}[c]{@{}l@{}}1. Discusses the opportunities of \\ UAV-enabled wireless communication\end{tabular} \\
 &  & \begin{tabular}[c]{@{}l@{}}2. Investigates the challenges of \\ UAV networks including, 3D \\ deployment, channel modeling, and\\ energy efficiency\end{tabular} \\
 &  & \begin{tabular}[c]{@{}l@{}}3. Review mathematical tools for \\ UAV applications\end{tabular} \\ \hline

 \multirow{3}{*}{\cite{hentati2020comprehensive}} & \multirow{3}{*}{\begin{tabular}[c]{@{}l@{}}UAV communication\\  protocols\end{tabular}} & 1. Discussion on UAV architectures \\
 &  & 2. Existing UAV communication protocols  \\ \hline
 \multirow{2}{*}{\cite{yan2019comprehensive}} & \multirow{2}{*}{\begin{tabular}[c]{@{}l@{}}Channel measurements \\ and modeling for UAV \\ and aeronautical \\ communications\end{tabular}} & \begin{tabular}[c]{@{}l@{}}1. Provides design guidelines for \\ managingthe link budget of UAV \\ communications\end{tabular} \\
 &  & \begin{tabular}[c]{@{}l@{}}2. Discussion on link losses and \\ channel fading effects\end{tabular} \\ \hline

\multirow{2}{*}{\cite{gupta2015survey}} & \multirow{2}{*}{\begin{tabular}[c]{@{}l@{}}UAV communication \\ networks\end{tabular}} & \begin{tabular}[c]{@{}l@{}}1. Discuss characterization of a\\ UAV  network\end{tabular} \\
 &  & \begin{tabular}[c]{@{}l@{}}2. Explore network set issues, \\mobility, and resource constraints \\of UAV network\end{tabular} \\ \hline
\multirow{2}{*}{\cite{fotouhi2019survey}} & \multirow{2}{*}{\begin{tabular}[c]{@{}l@{}}Commercial applications \\ of UAVs\end{tabular}} & \begin{tabular}[c]{@{}l@{}}1. Discussion on integration of \\ UAVs into cellular networks\end{tabular} \\
 &  & \begin{tabular}[c]{@{}l@{}}2. Interference issues and potential \\ solutions addressed by standard-\\ization bodies for providing \\aerial services\end{tabular} \\ \hline

\cite{nawaz2021uav} & \begin{tabular}[c]{@{}l@{}}Characteristics of a \\ UAV network\end{tabular} & \begin{tabular}[c]{@{}l@{}}1. Discusses the characteristics \\ of a UAV network compared to \\ WSN and mobile ad-hoc networks\end{tabular} \\
 &  & \begin{tabular}[c]{@{}l@{}}2. Summarizes the network \\ issues (such as routing, power\\ and quality of service) for \\ collaborative UAV networks\end{tabular} \\ \hline

\cite{alsamhi2021green} & \begin{tabular}[c]{@{}l@{}}UAV's role in realizing a \\ sustainable smart world\end{tabular} & \begin{tabular}[c]{@{}l@{}}1. Summarizes strategies to integrate \\ UAVs with IoT as edge intelligence \\ devices to collect and process data \\ obtained from IoT devices\end{tabular} \\
 &  & \begin{tabular}[c]{@{}l@{}}2. Explores the opportunities for \\ connectivity and communication \\ beyond 5G\end{tabular} \\ \hline
\cite{shi2021review} & \begin{tabular}[c]{@{}l@{}}UAV communication \\ protocols for power \\ line inspection industry\end{tabular} & \begin{tabular}[c]{@{}l@{}}1. Reviews the existing UAV \\ communication protocols for \\ power line inspection industry\end{tabular} \\
 &  & \begin{tabular}[c]{@{}l@{}}2. Classifies UAV communication \\ link type and summarizes wireless \\ mesh networking protocols\end{tabular} \\ \hline

\multirow{2}{*}{\cite{hayat2016survey}} & \multirow{2}{*}{UAVs for civil applications} & \begin{tabular}[c]{@{}l@{}}1. Discussion on the characteristics \\ and requirements of  UAV networks \\ for civil applications\end{tabular} \\
 &  & \begin{tabular}[c]{@{}l@{}}2. Investigate the suitability of \\ existing communication technologies \\ for UAV networks\end{tabular} \\ \hline

\multirow{2}{*}{\begin{tabular}[c]{@{}l@{}}Our \\ paper\end{tabular}} & \multirow{2}{*}{\begin{tabular}[c]{@{}l@{}}Collaborative communicati-\\ on and control mechanisms \\in UAVs\end{tabular}} & 1. Presents collaborative communication and control requirements for multi-UAV networks  \\
 &  & \begin{tabular}[c]{@{}l@{}}2. Reviews current collaborative \\communication mechanisms \end{tabular}
 \\
 &  & \begin{tabular}[c]{@{}l@{}}3. Presents UAVs applications for \\urban environments and smart cities \end{tabular}
 \\ \hline
\end{tabular}%
\end{table}

\subsection{Contributions and Organization}
We summarize the primary contributions of this review as follows:
\begin{itemize}
    \item First, we present the prospects of UAV collaboration and summarize their basic requirements and challenges, including communication, control, and cooperation.
    \item Then, we describe different collaborative tasking performed by a swarm of UAVs, such as joint task completion, trajectory formation, cooperative localization, data collection, and other cooperative decisions. In addition, we have also discussed the UAV swarm network applications in urban environments.
    \item After that, we present several use cases of collaborative UAVs to show their effectiveness. These use cases include agriculture and environment monitoring, remote sensing, surveillance, and disaster management.
    \item Finally, we highlight various exciting future research directions that can play a crucial role in advancing the potential of collaborative UAVs.
\end{itemize}

The rest of the paper is organized as follows: Section \ref{sectionII} thoroughly discusses the communication and control requirements and challenges of collaborative UAVs. Section \ref{sec:urbanenvir} presents UAV swarm network applications in urban environments. Section \ref{sectionIII} describes the use cases of UAVs applications. Section \ref{directions} highlights the future research directions, and finally, Section \ref{conclusion} summarizes the key findings of the paper.

\section{Collaboration in UAVs: An Overview} \label{sectionII}
At first, single UAV systems were used for navigation, surveillance, and disaster recovery, where each UAV works as an isolated node directly connected to the central ground station. However, in a single UAV system, due to independent operations of UAV in a designated area, they are more prone to system and communication failure. In addition, independent working of UAVs in a network also requires a longer time and a higher bandwidth to complete a mission. In contrast, in a multi-UAV system, UAVs work together to achieve a common objective. For example, UAVs can work collaboratively to generate high-resolution images and 3D mapping to identify hotspot areas during disaster relief. At the same time, the UAVs equipped with sniffers can detect a high level of methane to locate broken gas lines. Accordingly, UAVs can also supply water and food without endangering the lives of rescue personnel. Therefore, coordination and collaboration are crucial for achieving desired performance in a multi-UAV environment. This section will provide a detailed discussion on the requirements and challenges of collaborative UAVs, such as intelligence, communication, control, and collaboration. Furthermore, we will present state-of-the-art collaborative communication methods to highlight the UAV network's contributions and limitations.

\subsection{Intelligent UAVs}
The main components of a conventional UAV include sensing, communication, control, and a computational unit \cite{hentati2020comprehensive}. The sensing unit comprises multiple sensors integrated into a UAV for different purposes, such as assessing high-resolution objects,  temperature estimation, light detection, and antenna configuration \cite{adao2017hyperspectral}. At the same time, the communication unit enables UAVs to communicate with each other and with the central control station for exchanging information. A mandatory control unit generally controls the operations of conventional UAVs for collision avoidance, path planning, object tracking, and resource management. However, frequent communication with the central control unit and limited UAV-to-UAV communication capabilities limit the autonomy and collaboration of UAVs for independent mission completion. For example, in disaster relief operations, multiple UAVs with autonomous features can perform collaborative functions, such as a group of UAVs can examine the hazardous area while the other UAVs can perform supply-drop runs with medical aid to help the victims. In addition, UAVs with high-resolution cameras and resource management algorithms can perform intelligent decision-making to minimize damage. At the same time, when UAVs have an improved understanding of wind patterns in the urban environment, they can use that knowledge to avoid turbulence and choose minimal energy routes without taking frequent instructions from the control unit.

Similarly, intelligent UAVs with collaborative communication abilities can perform various distributed operations and make independent decisions for smart city applications \cite{alsamhi2021green}. For instance, UAVs working in the urban environment need a high level of coordination and collaboration with other sensing devices, machines, robots, drones, and people to perform certain operations. Accordingly, improved cooperation and knowledge of the deployment environment assist in the seamless interaction of UAVs with the surrounding objects for processing monitored data and making real-time decisions to enhance safety and reliability in complex environments \cite{geng2013mission}. Likewise, UAVs can improve object recognition by the semantic understanding of the surrounding objects in an urban environment to better understand and devise methods to interact with the surroundings.

However, the lack of efficient intelligent autonomous UAV-to-UAV communication mechanisms that enable independent UAV flight, trajectory formation, target localization and data manipulation decisions is hindering the scope of UAV-based applications. Thus, to benefit from the natural characteristics of UAVs (such as high mobility, flexible deployment, and different types of sensors integration), it is necessary to focus more on the autonomy and intelligent collaborative communication capabilities integration into UAVs to improve their performance as a team to understand the environment and share the knowledge and resources for intelligent decisions without a high dependency on the central control for systematic instructions \cite{chamoso2018use}.

\subsection{Communication Requirements}
In a multi-UAV system, UAVs communicate with each other and with the central backbone infrastructure to complete various designated tasks successfully. Communication among UAVs and infrastructure networks generally follows two modes of communication, UAV-to-infrastructure, and UAV-to-UAV, to exchange data and ensure a high level of connectivity to achieve collaborative communication. This section discusses the communication requirements of both modes in detail.
\subsubsection{UAV-to-Infrastructure}
\begin{figure*}[htb]
\center
\includegraphics[width=4.5in, height = 8cm]{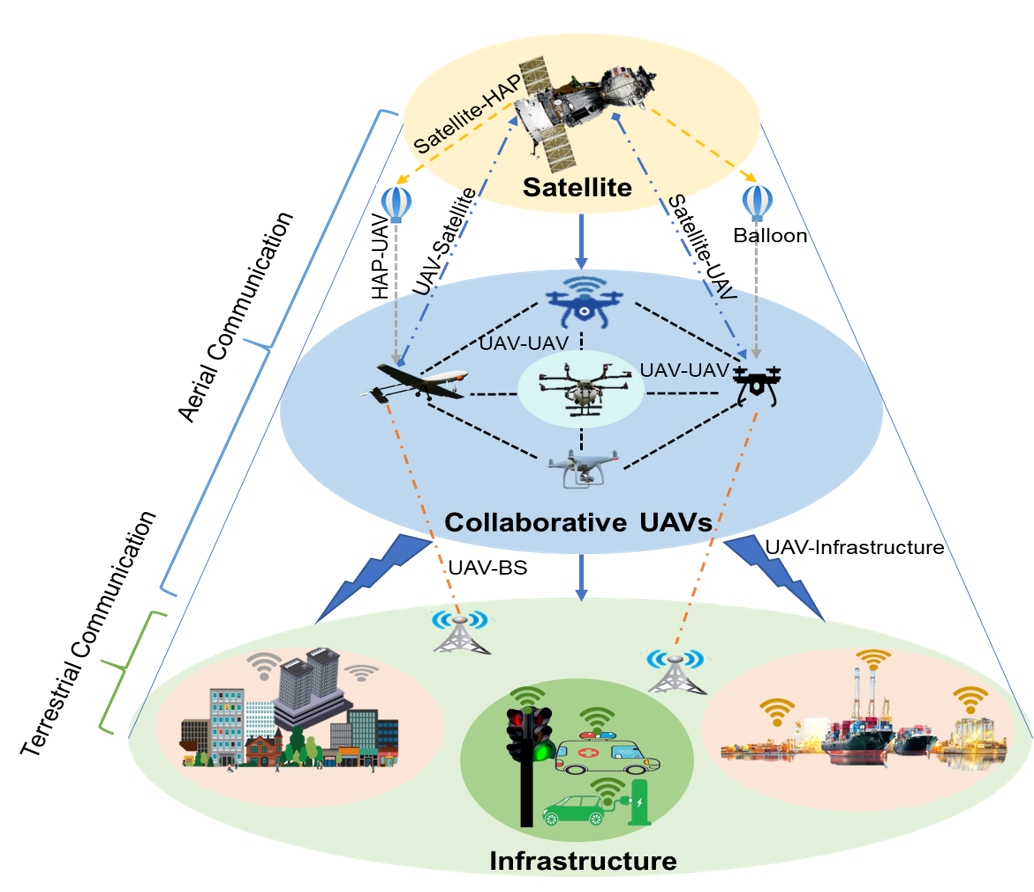}
\vspace{2.5 mm}
\caption{Overview of a communication links a collaborative UAV network.}
\label{Figurecolab}
\vspace{-4 mm}
\end{figure*}

UAV-to-Infrastructure communication enables information transfer between the UAVs and infrastructure network using different platforms, including terrestrial, high-altitude platforms (HAP), and satellites, as shown in Fig. \ref{Figurecolab}.  UAVs can act as a communication relay, user or a base station to establish effective communication. UAVs, as a relay, provide wireless coverage between ground stations and remote infrastructures when direct contact is unavailable. UAVs as a relay bring many advantages, such as improved coverage, fast speed, clear communication channel, easy deployment, and reliable data forwarding mode \cite{8493134, MAHMOUD2021331}. While UAVs can also work as a user to offload task to the edge server for enhanced coverage with low latency \cite{mcenroe2022survey}. At the same time, when UAVs work as base stations, they offer more flexibles solution to provide communication services as they enable better LoS propagation, scalability, and higher operational altitude for heterogeneous networks \cite{ 8658363 }. This section highlights the primary communication requirements for enabling an efficient swarm of UAVs using each platform mentioned above.

 \begin{itemize}
\item{UAV-to-Terrestrial:} It is the link between UAVs and on-ground Base Stations (BSs). These links can be established using LoS or Non-LoS communication links.  Nevertheless, the high altitude and increased mobility of UAVs raise various critical issues for UAV-to-Terrestrial communication, such as substantial aerial-terrestrial network interference, QoS requirements for UAV control messages, and LoS communication \cite{zeng2019accessing,wang2021hybrid}. In addition, UAV-to-Terrestrial communication is also affected by natural disasters such as forest fires, tsunamis, and earthquakes that destroy the communication entities and cause complete damage to backhaul networks \cite{grazia2014integration,zhang2016green}. To increase the data rate and reduce latency, recent works  \cite{li2018uav,zeng2019accessing} suggested integrating 5G and future cellular networks with the UAVs. In addition, satellite or direct communication links can overcome the terrestrial communication challenges in remote areas such as deserts, seas, and forests.
%Although the terrestrial systems can support Device to Device (D2D) communication using short link communication technologies for data dissemination, it is challenged by the limited availability of local resources such as power, spectrum, and cache \cite{fotouhi2019survey}.
\item{UAV-to-Satellites:} UAV-to-Satellite communication offers several benefits for connecting UAVs to the ground infrastructure compared to the UAV-to-Terrestrial communication, such as global coverage, improved navigation, and localization \cite{zeng2019accessing}. The advanced global coverage using low earth orbit satellites provides services in remote areas such as deserts and seas where terrestrial services are unavailable. Satellite links also provide beyond LoS communication, which is significantly beneficial for advancing global coverage and reliability of UAVs applications for tactical and commercial applications \cite{hosseini2019uav}. However, the high propagation loss and delays of UAV-to-Satellite communication links are significant challenges that need to be addressed for delay-sensitive applications \cite{marchese2019iot, saeed2020around}. Moreover, the operational costs for UAV-to-Satellite communication are also high, and the communication equipment, such as large antennas, consumes high energy \cite{zeng2019accessing}. In addition, during navigations, when UAV’s altitude varies frequently, they must constantly direct the beam towards the satellite to maintain the communication link \cite{zhao2018integrating}. Furthermore, effective channel modeling and physical layer characterization are also required for improved UAV-to-satellite communication  \cite{haas2002aeronautical,matolak2016air}.

\item{UAV-to-HAPs:} HAPs such as balloons and helikites are positioned highly elevated (up-to-20 km) compared to the terrestrial systems and offer multiple benefits, including broader coverage, advanced endurance, high data rate, and better path-loss characteristics. HAPs have low deployment costs compared to terrestrial setup, and UAV-to-HAP communication links also have better LoS communication that improves the signal reliability for direct communication while reducing the probability of LoS communication obstruction by the tall buildings in the urban environment \cite{zeng2016wireless,d2016high}. In addition, the HAPs platform is best suited to provide broadband communication services by carrying 3G, 4G, or beyond 4G payload that can play an essential role in providing emergency communication services in disaster areas \cite{morosi2013cooperative,dong2015energy}. HAPs have more advantages over terrestrial and satellite systems; however, spectrum regulation is the primary challenge for HAPs. HAPs need to address the global and regional spectrum challenges for frequency band allocation to enable flexible operation while protecting existing services within candidate frequency bands \cite{yuniarti2018regulatory}. Moreover, HAPs also raise safety, privacy, and security issues that need to be focused on providing secure UAV-to-HAP communication services \cite{he2016communication}.

\end{itemize}

%Different communication methods are introduced in the existing literature to support UAV-to-infrastructure and UAV-to-UAV links depending on the application requirements. For example, surveillance, security, and defense applications commonly use satellite links for long-term operations, better coverage, and reliability. On the other hand, civil and commercial applications prefer cellular communication technologies to alleviate various limitations such as range of communication, networking, and inadequate resources. {\color{green} Why are we suddenly talking about satellite and cellular after highlighting UAV-UAV and UAV-infrastructure links}.

\subsubsection{UAV-to-UAV Communications}
Besides UAV-to-Infrastructure communication, effective UAV-to-UAV links enable a swarm of UAVs to overcome various fundamental challenges such as autonomous flight, collision avoidance, distributed processing, and joint operations. The state-of-the-art literature has suggested multiple means to provide UAV-to-UAV communication, including satellite communication links, Wi-Fi links, Ultra-High Frequency (UHF) links, cellular communication links, Long-Range Wide-Area Network (LoRAWAN), and Free-Space Optical (FSO) links. This section discusses the requirements of these different approaches for enabling stable and reliable UAV-to-UAV communication.
\begin{itemize}

    \item{Wi-Fi and UHF radio links:} UAV-to-UAV communication using Wi-Fi radio links support communication only at a short-range and suffer from high interference. Wi-Fi radio links can provide BLoS communication using a ground control station as a correlated communication relay system. However, it increases delay and degrades the effectiveness of UAV-to-UAV communication for mission-critical applications. In contrast, UHF radio links provide better communication over long distances without getting affected by obstacles and diffraction. Point-to-point wireless link in the UHF band (400 MHz) also provides reliable communication in near-line or non-line of sight conditions. However, the spectrum regulatory bodies have not specified the frequency band for UHF radio links that are required for licensed protection \cite{becvar2017performance,militaru2018uav}.

    \item{Cellular networks-based links:} Civil and commercial applications prefer cellular communication technologies for UAV-to-UAV communication to alleviate various limitations such as range of communication, networking, and inadequate resources. Cellular-connected UAV also provides a cost-effective solution as they will reuse the millions of cellular BSs worldwide without the need to build new infrastructure dedicated to UAVs only \cite{zeng2019accessing}. However, cellular-connected UAVs have different communication and spectrum requirements that require new design considerations to avoid interference between the existing ground users and flying UAVs \cite{azari2020uav}. Moreover, BSs antennas conventionally provide 2D coverage; however, flying UAVs have high altitudes that require modifications in BS antenna design to provide 3D coverage \cite{zeng2018cellular}.

    \item{LoRaWAN-based links:} Besides cellular networks, LoRAWAN provides long-distance LoS communication for UAVs in a swarm, establishing UAV-to-UAV communication with high bandwidth and increased connectivity for exchanging a high amount of data \cite{xia2019emerging}. In addition, LoRAWAN extends the coverage area for different operational regions and enables low energy dissipation communication among UAVs \cite{khan2020implementation}. However, LoRAWAN operates at the unlicensed 900 MHz ISM band that experiences severe co-channel interference and path loss \cite{jouhari2022survey}. Therefore, interference mitigation approaches are required to provide LoRWAN-based reliable UAV-to-UAV communication \cite{behjati2021lora}.

    \item{FSO-based links:}
    Another alternative to establishing UAV-to-UAV links is FSO which uses light in the free space to transmit data wirelessly among UAVs \cite{fawaz2018uav}. FSO communication links are generally point-to-point or long-range links with high bandwidth for increased data rate connectivity, making them suitable for UAV-to-UAV communication to transmit a large amount of data wirelessly. Due to the long-range, the occasional loss of one of the UAVs operating in swarm formation does not entirely stop the data transfer process \cite{nallagonda2021performance}. Nevertheless, UAV’s high mobility is the primary constraint for establishing FSO-based links due to the misalignment between the transmitters and receivers. Moreover, atmospheric and turbulence effects also degrade the FSO-based UAV-to-UAV links, requiring effective channel and mobility modeling \cite{majumdar2014advanced}.

    \item{Satellite-based UAV-to-UAV communication links:} They provide secure LoS and BLoS communication to support stable and reliable UAV-to-UAV communication. The satellite-based UAV-to-UAV communication links are commonly used for surveillance, security, and defense applications for long-term operations, better coverage, and reliability. However, as UAVs are constantly moving, thus the high relative velocity results in a significant Doppler shift. In addition, satellite-based UAV-to-UAV communication links also suffer from  a high latency for time-critical applications \cite{saeed2020cubesat}. Therefore, more in-depth analyses are required to overcome the challenges of delay, Doppler shift, and multi-user interference when multiple UAVs are communicating using satellite-based UAV-to-UAV communication links \cite{hosseini2019uav}.

\end{itemize}

\subsection{Control Requirements}
Due to UAVs' small size and low-cost demand, they require a cost-effective control system that can enable flexible movement and trajectory tracking for takeoff, landing, hovering, maneuverability, altitude control, localization, and collision avoidance. The primary control requirements for UAVs are discussed below:

\begin{itemize}
\item {Landing and Takeoff:} The UAVs can be classified as fixed-wing and rotary-wing, where both have their specific requirements for landing and taking. The fixed-wing UAVs require a runway for takeoff and landing, while rotary-wing UAVs can take off and land vertically, improving their suitability for diverse civilian applications \cite{ rabelo2020landing,ghommam2017autonomous, saeed2021cars}.  In \cite{gu2017development}, the authors presented a hybrid Vertical Takeoff and Landing VTOL solution that integrates fixed-wing and rotary-wing UAV features in a single platform to achieve long flight endurance with high flight efficiency. The VTOL approach requires a fixed-wing position controller, rotary-wing position controller, transition controller, and VTOL mixer based on aerodynamic characteristics to realize transition and improved flight stability. In another work \cite{ ccakici2016control}, different PID controllers are used for VTOL without a runway and launch recovery equipment,  performing a smooth operation with control commands. Furthermore, in existing literature \cite{ kyristsis2016towards, araar2017vision, yang2016ground }, various solutions have also been presented that use visible light camera sensors, GPS, and IMU to land and take off using PID controllers.
    \item {Controlled Movement and Hovering:} UAV's rotors use propellers that enable roll, thrust control, pitch, yaw, and six degrees of freedom for spinning, maneuvering, and hovering. UAV's control algorithm adjusts the roll, pitch, and yaw for stable rotation on the X-axis, Y-axis, and Z-axis. Existing studies have presented various models to control the movement of UAVs; for example, Thu et al. in \cite {thu2017designing} modeled the well-known quadcopter control system for flexible movement and maneuver based on "+" and "×" flying configurations. In another work \cite{gheorghictua2015quadcopter}, a dynamic model is designed to control the UAV's motion on one rotation axis. Elkaim et al. in \cite{elkaim2015principles} presented a UAV control system that used position, velocity, and altitude estimation for UAV-controlled movement and trajectory formation. In another work in \cite{kwak2018autonomous}, an autonomous UAV flight control system is introduced that integrates a GPS to generate optimal flight paths. Furthermore, throttle movements, state information, and onboard sensing components have also been analyzed and modeled in existing literature for stable maneuver and hovering \cite{barton2012fundamentals,frank2007hover,azinheira2008hover}.

\item {In-flight Control}: UAVs state information such as the position and velocity are used to guide and control the UAV for precision operations such as landing or object tracking. Remotely Operated Aerial Model Autopilot (RAMA) is thoroughly described in \cite{ spinka2007control} that uses altitude, angular rates, and position information for designing the control system for small UAVs. Furthermore, PID controllers have also attracted significant attention from academia and industry for autonomous UAV operations \cite{ kada2011robust }. Integrated PID autopilot enables a complete set of avionics for autonomous UAV navigation and real-time operations \cite{noshahri2014pid}. In addition, the PID controller also improves reliability and stabilizes the movement of flying UAVs on a predefined trajectory with minimal error and energy consumption.

%\item {Localization} UAVs relays on GPS, inertial measurement unit (IMU) and various types of sensors and high definition cameras for accurate localization and tracking \cite{wu2017vision, xiao2018indoor}. Various localization approaches have been presented in the existing literature that integrated vision sensors and gas sensors for target tracking \cite{tisdale2008multiple, wolf2005robust,neumann2013gas}. The vision sensors are equipped with high-resolution cameras and require visual recognition algorithms and sensing capabilities to perform accurate vision-based sensing \cite{tang2020recognition}. On the other hand, gas localization requires integrating gas detection sensors. For instance, in \cite {neumann2013gas} gas-sensing devices such as metal-dioxide sensors and infrared sensors are incorporated into the micro-drone payload to detect toxic gases. In another work \cite{burgues2019smelling}, a smelling UAV is developed that integrates a custom sensing board to detect harmful gases. Similarly, in \cite{takei2019development} gas leakage is accurately detected by mounting gas sensors on the UAV's rotors to estimate the direction and concentration of the gas source.
\item {Collision avoidance} It is the basic requirement of UAV design to ensure autonomous UAV flight. Current literature \cite{ chee2013control,kwak2018autonomous} suggested various means such as GPS-guided navigation and different collision-avoidance sensors to avoid collisions. Furthermore, according to existing studies \cite{yasin2020unmanned, gageik2015obstacle }, cheap commercially available sensors (such as infrared, pressure, and height sensors) can easily be integrated into UAV flight systems to estimate the distance from an obstacle to control its movement. In addition, accurate location estimation of UAVs and trajectory planning is also a fundamental requirement for collision avoidance \cite{lin2017sampling}.
\end{itemize}

\subsection{Collaborative Tasking}
The existing developments regarding collaborative communication are focused comprehensively in this section. UAVs' collaborative tasking allows multiple UAVs to share information to perform various tasks cost and time effectively in a distributed manner with enhanced flexibility, robustness, and fault tolerance. In recent years, a few collaborative communication architectures have been presented, primarily focusing on integrating UAV networks with WSNs, Ad-hoc networks, and the IoT paradigm for effective monitoring and data collection \cite{ lin2020novel}. In addition, a few swarm-based approaches have also been introduced for collaborative trajectory planning, routing, and target localization.

\subsubsection{Swarm-based Collaborative Communication}
Several recent works use a swarm of UAVs to complete a mission collaboratively in a short time with better coverage, reliability, and efficiency. For example, Sathyan et al. in \cite{ sathyan2016efficient} presented a fuzzy genetic algorithm that is specific to polygon visiting multiple traveling salesman problems to solve the clustering and routing issues of UAVs swarms. The genetic fuzzy algorithm use distance for cluster formation and cost function optimization for efficient route selection to identify maximum targets with reduced computational complexity. In \cite{ sadrollah2014distributed}, the authors introduced a swarm of UAVs that are controlled and operated by a human operator at different levels to enable the self-organization of UAVs for monitoring and surveillance purposes. In another work \cite{de2015coordinating}, UAV’s collaborative movement and coordination are managed through mobile networks by integrating a smartphone into the UAVs. The UAVs fly in the region of interest either in a swarm or petrol mode to capture the images and transmit them to the ground station using the mobile network. Another work in \cite{chen2019uav} presented a joint task placement and routing algorithm for a UAV swarm network. In the UAV swarm network, each UAV is assigned a different task and path by the central algorithm to complete the collaborative operation with reduced latency, improving QoS demand. In \cite{arafat2019localization}, the particle swarm optimization algorithm is presented for optimal localization and clustering of the swarm of UAVs. Once targets are identified, the swarm intelligence-based algorithm forms the cluster for multi-hop communication and retrieves the emergency information with reduced energy consumption.

\subsubsection{Deep Reinforcement Learning for UAV swarm networks}
In the past few years, reinforcement learning techniques have been considerably adopted to improve the performance of UAV swarm networks for path planning, navigation, and control in complex environments \cite{koch2019reinforcement,azar2021drone}. For example, in \cite{xia2021multi}, a multiagent reinforcement learning technique is introduced for UAV Swarm-based target tracking. End-to-end cooperative multiagent reinforcement learning allows UAVs to make intelligent flight decisions based on the past and present states of the target for cooperative target tracking with reduced energy consumption. In another work \cite{hu2022autonomous}, Hu et al. proposed an autonomous maneuver decision-making scheme based on deep reinforcement learning for cooperative air combat. UAVs perform the situation assessment to identify the target's current situation and design the reward function for increasing the training convergence of UAVs for defeating the enemy. In \cite{wang2019autonomous}, authors presented a deep reinforcement learning method that allows UAVs to execute navigation tasks in large-scale complex environments for remote surveillance and goods delivery application. Reinforcement learning-based navigation enables direct mapping of UAVs' raw sensory measurements into control signals for autonomous navigation in large-scale, complex, and three-dimensional environments.

Furthermore, Zhang et al. in \cite{zhang2015geometric} presented a geometric reinforcement learning scheme for UAV's path planning by exploiting a specific reward matrix to select the candidate points from the current position to the target position for effective navigation. In another work \cite{ challita2019interference }, authors designed an interference-aware path planning scheme for cellular-connected UAVs based on an echo state network deep reinforcement learning algorithm. The deep echo state network architecture enables each UAV to map each observation of the network state to action for reducing a sequence of time-dependent utility functions and learn about its transmission path, optimal path, and cell association vector for minimizing the interference level and transmission delay. In \cite{ liu2018energy}, the authors introduced a deep reinforcement learning algorithm based on deep neural networks for UAV control. Deep neural network-based reinforcement learning algorithm leads to significantly enhanced energy efficiency, coverage, and connectivity of the UAV swarm network.

In another work \cite{hu2020reinforcement},  Hu et al. focused on improving the sensing and communication quality of cellular-connected UAVs by designing a reinforcement learning algorithm-based distributed sense-and-send protocol. The reinforcement learning algorithm-based distributed protocol enhanced UAV coordination and led to efficient collaborative trajectory control and resource management. Cui et al. in \cite{cui2019multi} also improve resource allocation in the UAV swarm network by introducing a multiagent reinforcement learning framework that allows each UAV to work as a learning agent. Each action of UAVs as a learning agent corresponds to a resource allocation solution and helps them to find the best solution for the resource allocation based on their local observations. In \cite{ zhang2020hierarchical}, the authors introduced deep reinforcement learning for backscattering data collection with multiple UAVs to reduce their total flight time. The deep reinforcement learning algorithm allows UAVs to fly within the deterministic boundary and enable cooperative learning to find the ambiguous backscatter sensor node for data collection. In addition, in \cite{qu2020novel}, the deep reinforcement learning algorithm has also been used for three-dimensional complex flight environments by introducing a reinforcement learning-based grey wolf optimizer algorithm for UAV path planning. The grey wolf optimizer algorithms mimic the social hunting behavior of the grey wolf and perform exploration, exploitation, geometric adjustment, and optimal adjustment for smooth planning of the UAV's flight route.

\subsubsection{Trajectory Formation}
\begin{figure}[t]
\center
\includegraphics[width=3in, height = 5cm]{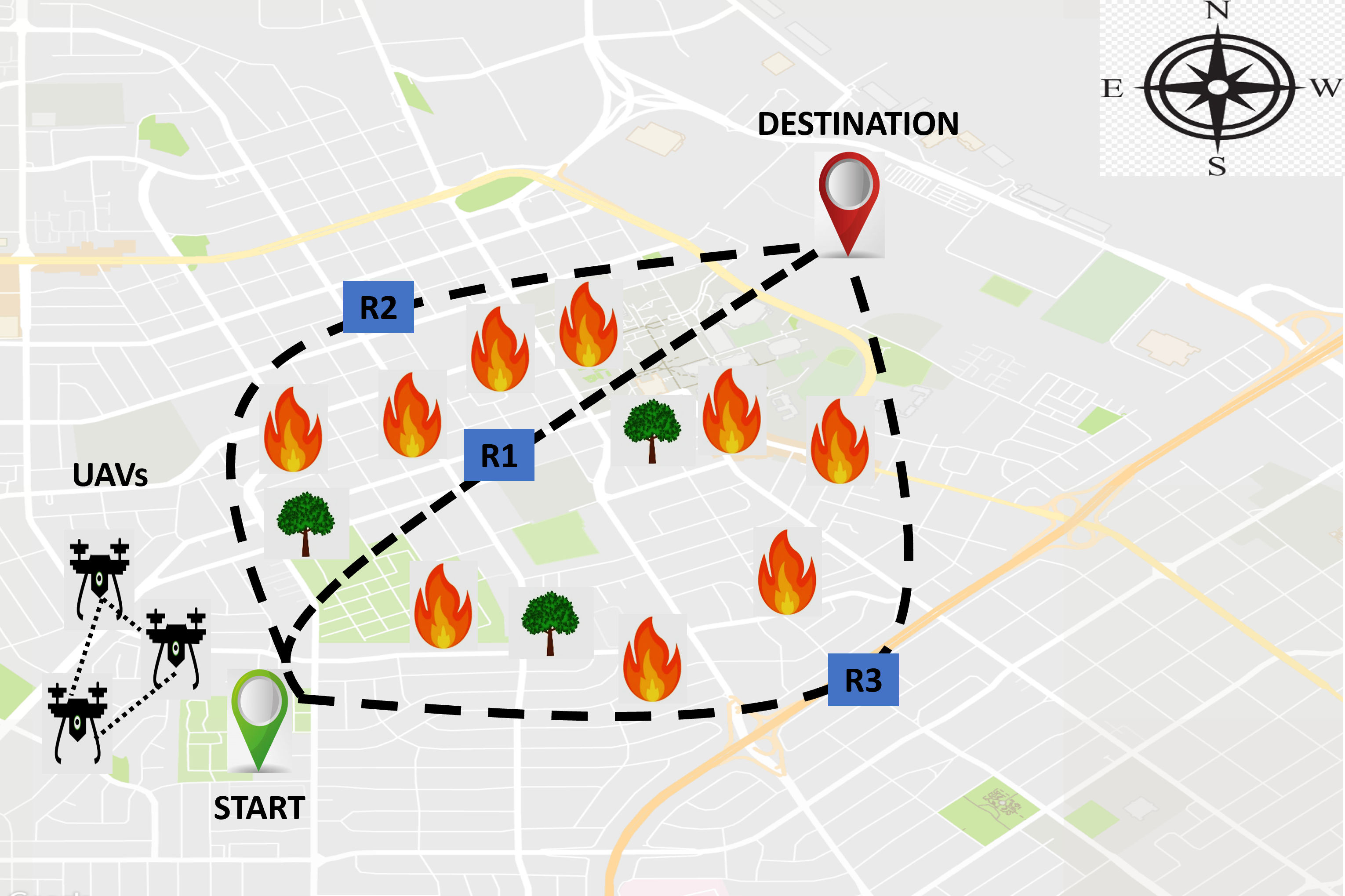}
\vspace{2.5 mm}
\caption{Trajectory optimization using collaborative UAVs.}
\label{Figuretraj}
\vspace{-4 mm}
\end{figure}

Collaborative trajectory formation enables multiple UAVs to find the optimal path from the starting point to the target point. It is one of the emerging areas of research in UAV systems as collaborative path planning minimizes localization cost, improves maneuver decisions, and helps in collision avoidance \cite{ Imran2022, schouwenaars2004decentralized, pr10071236}. In the existing literature, various collaborative trajectory formation techniques exist. For example, in \cite{wei2018ucav}, a dynamic trajectory planning method is introduced, in which the leader UAV uses the hp-adaptive pseudospectral method to find the optimal path and shares it with the follower UAV for operation execution. The hp-adaptive pseudospectral method begins with a global pseudospectral approximation for the state. Then, each iteration determines locations for the segment breaks and the polynomial degree for the next iteration to select the optimal path in a dynamic environment in a reduced time. In another work in \cite{ ji2019fair}, optimal path planning is performed by solving multiple traveling salesman problems using the genetic algorithm. Each UAV flies randomly to locate the target and share this information with the central BS. The central BS executes the information collected from the flying UAVs and estimates the target location for trajectory formation with reduced complexity and energy consumption. Wang et al. in \cite{wang2017minimum} introduced the Sequential Convex Programming (SCP) method for cooperative trajectory formation that iteratively finds an optimal local solution with reduced computational complexity. Therefore, it has been extensively used for trajectory planning with highly nonlinear dynamics, such as proximity operations, spacecraft rendezvous, and fuel-optimal powered landing \cite{liu2016entry,szmuk2016successive}.
In  \cite{ tang2019optimized}, the optimized Artificial Potential Field (APF) method is proposed for UAVs collaborative path formation. UAVs use optimized APF (APF was originally introduced for moving robots from initial points to the targeted points \cite{khatib1986real}) to perform dynamic steps adjustment and climb strategy for safe and stable paths selection for their movement with reduced collision probability.
In \cite{guo2019ultra}, Guo et al. introduced a cooperative relative localization scheme to enable infrastructure-free communication among UAVs for UAVs localization and flight formation using consensus-based fusion. Each UAV performs consensus-based fusion for direct and indirect estimation of a target location and uses this information to control flight trajectory. Recently, Qadir et al. \cite{9659824} compared different metaheuristic algorithm for UAVs efficient path planning in complex environments. The autonomous trajectory optimization for UAV shows promising results in naturally occurring catastrophic events as shown in Figure \ref{Figuretraj}.

\subsubsection{Cooperative Target Localization}
Accurate localization brings significant advantages for target indication, aerial filming, data sensing, and air-to-ground attacks \cite{THzCommag20}. Collaborative UAVs communication helps identify the target in reduced time with better position accuracy \cite{KHALIL2022}. Therefore, various methods have been introduced in the literature to enable collaborative UAVs localization with reduced communication delay and packet loss. For instance, in \cite{ fu2017multi}, authors introduced a centralized and distributed mechanism for cooperative UAVs target localization with reduced communication delay and packet loss. Each follower UAV separately transmits the target information to the leader UAV during centralized localization without any prior processing. The leader UAV processes all obtained information for the final target location estimation.
On the other hand, in disturbed target identification, multiple UAVs cooperatively process the location information and transfer the final positioning information to the leader UAV for further processing. In  \cite{shahidian2016optimal}, Received Signal Strength (RSS) and Differential RSS (DRSS) are used to identify RF sources and plan UAVs' movement toward the RF targets. Then, in \cite {fu2019pollution }, Fu et al. improved pollution source localization by enhancing the convergence and the searchability of particles by using APF and particle swarm optimization for cooperative UAV communication.

Furthermore, in \cite{lee2013cooperative}, a cooperative maneuver scheme is introduced in which two UAVs are equipped with heterogeneous sensors (i.e., bearing-only and range-only sensors) for cooperative localization. The use of heterogeneous sensors for cooperative maneuver localization led to reduced complexity and enhanced data collection in a shorter time. Another work \cite{cui2015drones} simulates and demonstrates a cooperative search and rescue operation for a post-disaster situation using multiple UAVs. Multiple UAVs perform different tasks such as localization, inspection, path planning, and navigation to identify the houses and survivors in the targeted area for collaborative search and rescue operations.

\subsubsection{Data Collection}
In the past few years, numerous collaborative communication mechanisms have been introduced to improve the remote data collection experience \cite{mahmoud2014collaborative}. Furthermore, various existing schemes have also integrated UAV networks with WSN and IoT to enhance their collaborative performance. For example, in \cite{popescu2018collaborative}, a WSN-UAV collaborative network is presented to optimize data collection from a resource constraint sensor network by introducing a UAV that follows a predefined trajectory for energy-efficient and fast data acquisition. Similarly, in \cite{wang2020minimizing}, multiple flying UAVs collaboratively collect data from the sensing devices in a short time with better efficiency and reduced energy consumption. In \cite{luo2020revenue}, UAV and fog computing-based data communication architecture (consisting of multiple UAVs, fog nodes, and ground station) is presented to collect data from natural disaster and hazard monitoring areas efficiently. During an emergency, when flying UAVs cannot transfer data to the ground station directly, then the flying UAVs transfer the collected data to the fog nodes for storage, processing, and transmission to the ground station for effective remote processing and rescue operations. Wang et al. in \cite{wang2020multi} also introduced multiple flying UAVs for data collection from IoT devices. The flying UAVs follow their predefined path to collect data from their designated areas with the aim of equal opportunity for all UAVs to participate in data aggregation with a reduced flight time to save energy.

\begin{figure}[t]
\center
\includegraphics[width=3in, height = 5cm]{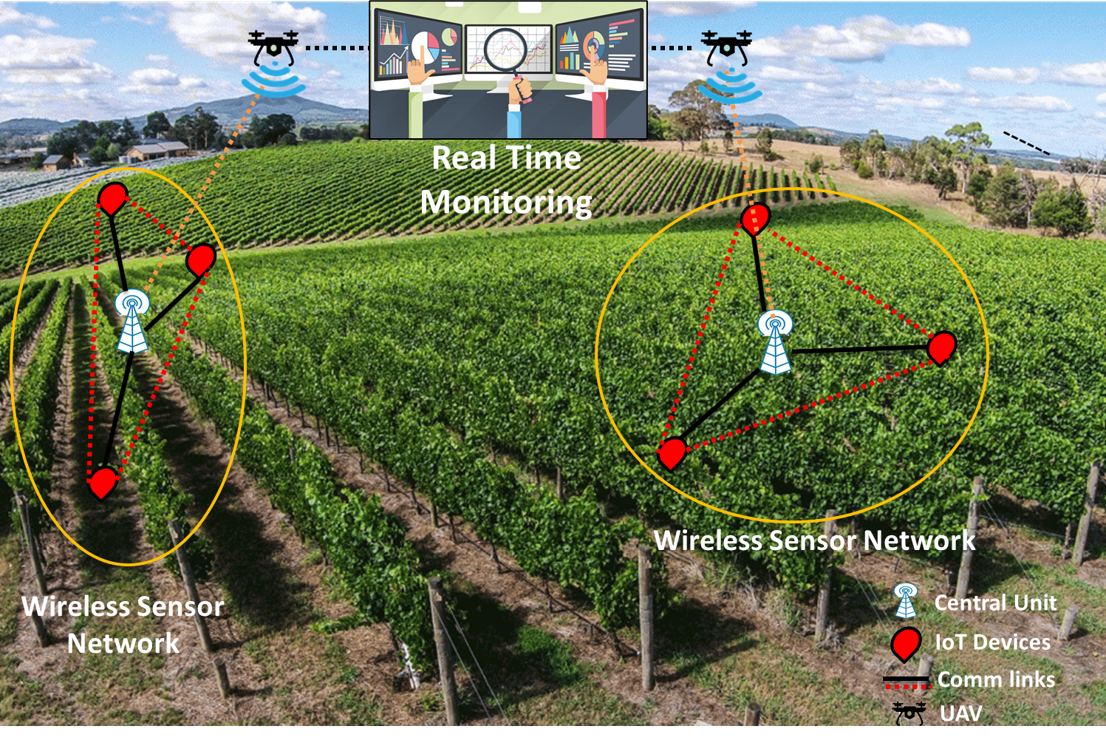}
\vspace{2.5 mm}
\caption{Collaborative UAV-WSN-IoT in an agricultural field.}
\label{agri}
\vspace{-4 mm}
\end{figure}

In \cite{popescu2020advanced}, a collaborative UAV-WSN-IoT communication architecture is presented comprising multiple UAVs, sensing, and IoT devices deployed in an agriculture field
as shown in Figure \ref{agri}. UAVs search for IoT devices to efficiently collect the monitored data energy for precision agriculture. In \cite{kumar2020collaborative}, UAV-WSN-IoT is introduced for collaborative data acquisition and processing for post-disaster management using cloud services. Multiple UAVs fly in their predefined clusters to localize sensing and IoT devices. The collected data is stored in the cloud for efficient data processing and management. Schmuck in \cite{schmuck2017multi} presented a new collaborative UAV communication architecture based on the Simultaneous Localization And Mapping (SLAM) algorithm where multiple flying UAVs run the SLAM algorithm to enable independent exploration of their environment. Each UAV communicates its experience with the ground station, and then the ground station uses the collected information for data fusion, advanced processing, task assignments, management, and optimization. In \cite{liu2020opportunistic}, the authors introduced UAVs for opportunistic data collection in cognitive WSN where the ground sensing device forms clusters using the coalition formation game model. Coalition game theory allows ground sensing devices to share resources and balance the load to upload the data to the flying UAVs with improved reliability and efficiency.

\subsubsection{Cooperative Decisions}
The autonomous operations of UAVs require making complex decisions to meet the application objectives, such as the elimination of threats or time-critical rescue operations. These complex decisions are affected by inadequate information, high uncertainty, delay, and task coupling \cite{shima2009uav}. Therefore, UAVs need cooperative decision algorithms to provide robustness and flexibility to ensure adequate performance in uncertain and adversarial environments. Zhao et al. in \cite{zhao2019systemic} introduced the max-consensus algorithm for estimating the Joint Multi-Target Probability Distribution (JMTPD) of UAVs for cooperative multi-target identification decisions. The max-consensus protocol employs a distributed information fusion strategy to improve the JMTPD estimation and lead to accurate observation and estimation of the moving targets. In \cite{capitan2016cooperative}, Partially Observable Markov Decision Processes (POMDPs) are adopted for cooperative surveillance decisions, which are modeled for different surveillance problems, and each UAV is assigned a different behavior to complete the surveillance policy cooperatively. Furthermore, various state-of-the-art schemes have focused on cooperative decisions for task assignments, control, and coordination. However, the joint decisions of UAVs need more attention for reducing delay and uncertainties in dynamic environments for stable operations of multiple UAV systems. The state-of-the-art
developments regarding collaborative communication are also further summarized in Table \ref{colltasks}.
%\begin{table*}[]
%\caption{State-of-the-art Collaborative Tasks Mechanisms}
%\label{colltasks}
%\renewcommand{\arraystretch}{1.5}
%\resizebox{\textwidth}{!}{%

% \usepackage{graphicx}
\begin{table*}[]
\caption{State-of-the-art works on collaborative tasks mechanisms.}
\label{colltasks}
\renewcommand{\arraystretch}{1.5}
\resizebox{1.08\textwidth}{!}{%
\begin{tabular}{llllll}
\hline
\textbf{Ref.} & \textbf{Purpose} & \textbf{Advantages} & \textbf{Ref.} & \textbf{Purpose} & \textbf{Advantages} \\ \hline
\cite{wei2018ucav} & \begin{tabular}[c]{@{}l@{}}Optimal trajectory formation in a \\ dynamic environment\end{tabular} & \begin{tabular}[c]{@{}l@{}}1. A new trajectory can be\\ formed at low cost when \\ a new event occurs\end{tabular} & \cite{ji2019fair} & \begin{tabular}[c]{@{}l@{}}Energy-efficient cooperative trajectory \\ formation using genetic algorithm\end{tabular} & \begin{tabular}[c]{@{}l@{}}1. Low computational complexity \\ 2. Energy conservation\end{tabular} \\ \hline
\cite{wang2017minimum} & \begin{tabular}[c]{@{}l@{}}SCP-based cooperative trajectory \\ formation for reduced computational \\ complexity\end{tabular} & \begin{tabular}[c]{@{}l@{}}1. Minimum time trajectory \\ formation  \\ 2. Low computational complexity\end{tabular} & \cite{tang2019optimized} & \begin{tabular}[c]{@{}l@{}}Optimal path planning and collision \\ avoidance for stable path formation\end{tabular} & \begin{tabular}[c]{@{}l@{}}1. Stable trajectory formation\\ 2. Collision avoidance\end{tabular} \\ \hline
\cite{cui2015drones} & \begin{tabular}[c]{@{}l@{}}Cooperative search and rescue in a \\ post-disaster situation for localization, \\ inspection, path planning and navigation\\ to identify the houses and survivors \\ in the targeted area\end{tabular} & \begin{tabular}[c]{@{}l@{}}1. Collaborative search and rescue\\ operations simulation and \\ demonstration\end{tabular} & \cite{guo2019ultra} & \begin{tabular}[c]{@{}l@{}}Consensus-based location estimation \\ and formation control for target \\ identification\end{tabular} & \begin{tabular}[c]{@{}l@{}}1. UAVs formation without \\ infrastructures, global positions, \\ and pattern detection \\ computation\end{tabular} \\ \hline
\cite{fu2017multi} & \begin{tabular}[c]{@{}l@{}}Centralized and disturbed localization  \\ with precision and accuracy\end{tabular} & \begin{tabular}[c]{@{}l@{}}1. Reduced communication delay \\ and packet loss \\ 2. Improved position estimation\end{tabular} & \cite{shahidian2016optimal} & \begin{tabular}[c]{@{}l@{}}RSS and DRSS are used for trajectory\\ planning and location estimation\end{tabular} & \begin{tabular}[c]{@{}l@{}}1. Minimized number of UAVs \\ 2. Reduced mission time and \\ path length\end{tabular} \\ \hline
\cite{fu2019pollution} & \begin{tabular}[c]{@{}l@{}}Cooperative location estimation for\\ identifying pollution sources\end{tabular} & \begin{tabular}[c]{@{}l@{}}1. APF helps in collision avoidance \\ 2. Reduced cost and efficient target\\ localization\end{tabular} & \cite{lee2013cooperative} & \begin{tabular}[c]{@{}l@{}}Cooperative maneuver localization \\ using two small UAVs that are equipped\\ with heterogeneous sensors\end{tabular} & \begin{tabular}[c]{@{}l@{}}1. Low complexity\\ 2. Enhanced data collection in\\ a limited time\\ 3. Improved accuracy\end{tabular} \\ \hline
\cite{wang2020minimizing} & \begin{tabular}[c]{@{}l@{}}Collaborative data collection in reduced\\ time\end{tabular} & \begin{tabular}[c]{@{}l@{}}1. Minimized data collection time\\ 2. Reduced energy consumption\end{tabular} & \cite{luo2020revenue} & \begin{tabular}[c]{@{}l@{}}UAV-fog collaborative communication\\ architecture for remote data collection\end{tabular} & \begin{tabular}[c]{@{}l@{}}1. Improves the revenue of \\ UAV-enabled remote data\\ collection\end{tabular} \\ \hline
\cite{wang2020multi} & \begin{tabular}[c]{@{}l@{}}Multiple flying UAVs follow their\\ predefined path to efficiently collect data \\ in a reduced flight time\end{tabular} & \begin{tabular}[c]{@{}l@{}}1. Reduced flight time\\ 2. Low energy consumption\end{tabular} & \cite{popescu2020advanced} & \begin{tabular}[c]{@{}l@{}}UAVs search for sensing and IoT devices \\ to efficiently collect data for precision \\ agriculture\end{tabular} & \begin{tabular}[c]{@{}l@{}}1. Energy efficiency \\ 2. High precision\end{tabular} \\ \hline
\cite{kumar2020collaborative} & \begin{tabular}[c]{@{}l@{}}Multiple UAVs fly in their predefined \\ clusters to localize sensing and IoT devices\\ for efficient data collection\end{tabular} & \begin{tabular}[c]{@{}l@{}}1. Energy efficiency\\ 2. Improved data management\end{tabular} & \cite{schmuck2017multi} & \begin{tabular}[c]{@{}l@{}}Localization and data collection using \\ SLAM algorithm for independent \\ exploration of the environment\end{tabular} & \begin{tabular}[c]{@{}l@{}}1. Energy efficiency \\ 2. Optimized resource utilization\end{tabular} \\ \hline
\cite{liu2020opportunistic} & \begin{tabular}[c]{@{}l@{}}Coalition game theory is used that allows\\ ground sensing devices to share resources\\ and balance the load for opportunistic \\ data transmission\end{tabular} & 1. Improved reliability and efficiency & \cite{zhao2019systemic} & \begin{tabular}[c]{@{}l@{}}The cooperative decision for multi-target\\ tracking\end{tabular} & \begin{tabular}[c]{@{}l@{}}1. Improved observation\\ and  location estimation\end{tabular} \\ \hline
\cite{capitan2016cooperative} & The cooperative surveillance decision & \begin{tabular}[c]{@{}l@{}}1. Prove the POMDPs for cooperative\\  surveillance decisions\end{tabular} & \cite{ben2008distributed} & \begin{tabular}[c]{@{}l@{}}Cooperative control and decision using \\ Ad-hoc communication for tasks assignment \\ and coordination\end{tabular} & \begin{tabular}[c]{@{}l@{}}1. Effective bandwidth \\ utilization \\ 2. Accurate target localization\end{tabular} \\ \hline
\end{tabular}%
}
\end{table*}

\section{UAV Swarm Network Applications in Urban Environments} \label{sec:urbanenvir}
UAVs are an integral part of sustainable urban development due to their crucial roles in accelerating the development of various smart city applications such as transportation, surveillance, security, key infrastructure monitoring,  networking and disaster relief \cite{qadir2021addressing,menouar2017uav,kuru2021planning}. Accordingly, the current literature substantially focuses on UAVs for the seamless integration of information and communication technology for realizing the concept of a smart city for urban development. This section summarizes the developmental challenges and state-of-the-art solutions of UAV swarm networks in different smart city applications.
\subsection{Intelligent transportation system} An intelligent transportation system is one of the primary components of smart city applications. However, the high mobility of vehicles, the large number of obstacles, bridges, and tall buildings lead to substantial connectivity and communication issues for effective management of transportation systems in an urban environment. Therefore, UAV-aided transportation systems are considered one of the optimal solutions to ensure the essential requirement of high connectivity and automation of vehicles and other building blocks of the smart transportation system (such as traffic police, road surveys, and rescue teams) \cite{xiong2012intelligent, brahim2014roadside, zhang2008uav}. UAV-aided transportation systems offer increased mobility and fast response in an emergency. In addition, a UAV swarm network can also optimize vehicle route selection to avoid congestion and help better enforce traffic rules and regulations \cite{ menouar2017uav}. State-of-the-art literature \cite{fawaz2018effect, khan2017uav} have introduced UAV-based vehicular networks to use a swarm of UAVs to work collaboratively to provide increased coverage, robustness and connectivity for urban vehicular networks. In \cite{oubbati2019uav}, the authors introduced a hybrid model based on vehicular ad hoc networks and UAV swarm networks to enable communication between vehicles and UAVs. The hybrid framework focus on finding a reliable routing path and tracking the expiration of a routing path to ensure a high level of connectivity for a stable transportation system. The authors in \cite{oubbati2017intelligent} also introduced UAV-assisted routing to enable fast data delivery for urban vehicular networks. However, the collaborative operations of UAV swarm networks need more efforts to facilitate effective UAV-to-UAV, UAV-to-vehicle and UAV-to-infrastructure communication for the practical realization of an intelligent transportation system.

\subsection{Intelligent environmental monitoring systems} The realization of the smart city require intelligent monitoring systems to manage its resources and infrastructure for better connectivity and services. UAVs with different types of integrated sensors and communication equipment present an excellent solution to monitor a large area independently. UAVs can perform remote sensing and provide high-resolution images, video footage, and multispectral, hyperspectral, and thermal imagery that helps them work as independent monitoring stations to execute various operations when required \cite{green2019using,jumaah2021development}. However, due to the natural challenges of monitoring a complex and large area, environmental monitoring requires a UAV swarm network to work cooperatively with the ground sensing devices to monitor a large area and share the resources for efficient data manipulation and decision. In addition, the high-level collaboration among UAVs and ground sensing devices also needs to reduce latency and enable real-time communication to achieve high QoS requirements for intelligent monitoring applications.

Integrating UAV swarm networks with the IoT paradigm offers advanced monitoring opportunities by connecting ground sensing devices with flying UAVs to address the above-mentioned issues \cite{alsamhi2021green}. Various existing UAV-IoT frameworks \cite{ hernandez2018internet, popescu2020advanced, kumar2020collaborative} focus on connecting the UAV swarm network with IoT architecture to enable continuous monitoring. In \cite{ mohamed2017uavfog}, authors introduced UAV-fog to reduce latency and improve scalability and real-time communication for UAV-IoT-based applications. In another work \cite{ madridano2021software}, the authors presented a software architecture framework for autonomous operations of a UAV swarm network to carry out coordinated firefighting operations in dense urban or forest environments. The software architecture includes a set of complementary methods that allow UAVs to communicate autonomously for cooperative navigation and establish a scalable, robust, and secure communication system for firefighting in dense environments. Moreover, in \cite{ sharma2022uav}, Sharma et al. also introduced the UAV-IoT framework to monitor the air quality for detecting toxic gases and finding the cause of air pollution to mitigate its effect on human health. However, the massive potential of UAV-based environmental monitoring requires more attention toward developing advanced UAV collaborative networks to ensure a high level of connectivity and networking to enable cost-effective long-term intelligent monitoring.

\subsection{Intelligent surveillance} Security and safety have always been the foremost concerns for urbanization. Accordingly, realizing the concept of the smart city is impossible without an intelligent surveillance system that can provide safety and security to the residents with a well-established secure infrastructure. The deployment of UAVs with integrated high-definition cameras can be a security measure to track intruders and monitor unsafe activities such as violence and theft. In addition, a swarm of UAVs can also work together to provide scalable surveillance in a large area with advanced recognition and detection abilities to prevent crimes and to provide safety and security to the citizens around the clock. Various UAV-based surveillance applications for police forces, emergency responders, and industrial and border inspection and security personnel have already been successfully implemented.

Consequently, state-of-the-art literature \cite{giyenko2016intelligent,cooley2018secure, jain2021towards} focuses on developing UAV-based intelligent surveillance systems for urban environments. For example, Liu et al. in \cite{ liu2020reinforcement} introduced a reinforcement learning-based control framework for a swarm of UAVs to perform persistent cooperative surveillance in an unknown urban area. UAV maneuver and target localization using an artificial neural network improve target identification and lead to quality surveillance in the complex urban environment. In another work \cite{ jin2020uav}, the authors introduced a UAV-based surveillance network that optimizes the performance of smart devices with integrated cameras to enable reliable and real-time large-scale video surveillance with reduced latency and improved throughput and QoS. Furthermore, current literature \cite{ullah2020applications,thakur2021artificial} also emphasizes using artificial intelligence techniques to provide cost-effective optimal surveillance quality.

\subsection{Edge computing based resource management} UAV's diverse operation for providing surveillance, internet coverage, rescue and relief demands long-term real-time monitoring leading to the extensive amount of video data generation that needs to be communicated and managed \cite{ishtiaq2021edge}. The UAV-based system needs to perform video analytics operations to digitally visualize the video inputs through advanced machine learning algorithms to transform video data into intelligent data for smart decision-making. However, real-time data monitoring and management require addressing several crucial challenges, such as QoS degradation, communication latency, and inadequate computational and storage capacity. Edge computing brings opportunities to solve these shortcomings by providing services at the network's edge to avoid data transmission to remote servers for data manipulation and decisions. Accordingly, various edge-computing-based schemes have been introduced for resource management of smart city applications \cite{zheng2020accurate,yang2019energy, jeong2017mobile,zhou2018uav,javaid2018traffic,javaid2022medical}. For example, Chen et al. in \cite{chen2019uav} introduced a hybrid model called UAV-Edge-Cloud to improve the QoS and resource provision for resource-intensive applications for smart city applications. The UAV-Edge-Cloud framework consists of a UAV swarm layer that forms an edge layer near the UAV swarm to enable fast real-time interaction with the users with better quality and resource utilization. Another work \cite{li2019near} introduces energy-efficient rechargeable UAV deployment to optimize the energy efficiency of the UAVswarm network for improved seamless radio coverage in an urban environment.

In another work \cite{zhan2020completion}, the UAV-enabled mobile edge computing system is introduced to reduce energy consumption. UAVs fly near the IoT devices to complete different tasks with local computational capabilities for efficient resource management and minimization of energy resources of UAVs and IoT devices. In \cite{zhang2019joint}, the authors presented a joint computation and communication architecture for IoT and UAV-assisted mobile edge computing framework. UAVs with integrated edge computing server handles the local data of the IoT devices to reduce latency and save communication and computation energy of the network. Furthermore, Yang et al. in \cite{ yang2020multi} also used a swarm of UAVs with integrated edge computing servers for offloading tasks of IoT devices to improve QoS and network performance. The above discussion shows that UAV-enabled edge computing services significantly improve the QoS provisioning and resources management of various WSN and IoT architectures. However, more efforts are required to optimize the communication among IoT devices and UAVs for better task offloading and resource sharing.

\section{Use Cases of Collaborative UAVs} \label{sectionIII}
Collaborative UAVs have already become a crucial part of various real-world applications. This section presents some of the use cases of UAVs to highlight their role in advancing various industries.
\subsection{ Agriculture and Environmental Monitoring}
UAVs continue to be widespread in agricultural, forest, and environmental monitoring. In agriculture, UAV technology plays a significant role in improving crop production to manage the rising global population pressure on agriculture consumption. For example, UAV-based soil information sourcing is beneficial at the early stage of crop cycles for soil analysis, seed planting patterns scheduling, and irrigation planning to determine the precise quantity of fertilizer needed for improved yield quantity. Agricultural UAVs also assist farmers in crop spraying by covering a large area in a short amount of time with high precision. UAV-based efficient spraying can reach both the plants and the soil below and can protect the farmers against prolonged exposure to potentially dangerous chemicals that have previously been linked with manual spraying \cite{ yinka2019sky}. Furthermore, UAV technology can enable extensive forest monitoring to support policies and decisions to conserve, protect, and sustainably manage forests. UAV-based forest monitoring systems can consistently assess forest cover and carbon stock change, especially in the tropics where forests are rapidly vanishing \cite{ dainelli2021recent}. Recently, drones, satellites, and mobile phone apps have already been used to protect the forest from deforestation in the Peruvian Amazon \cite{WinNT}. Moreover, UAVs can also plant trees in the forest at a precise location by analyzing the soil and existing plants.

The swarm of UAVs has the potential to perform all the above-discussed tasks in a more cost and time-effective manner. For example, a swarm of UAVs can work with the ground IoT paradigm deployed in an agriculture field to aggregate crop condition monitoring data and cover a large field in a relatively more shorter time. In addition, UAVs equipped with different features such as high-definition cameras, sprays, and edge computing servers can collect detailed information, perform data processing operations with intelligent algorithms and make decisions such as spraying the water or fertilizers to improve crop conditions.

\subsection{Advanced Surveillance} Advanced surveillance applications require long-term monitoring of the interested area with high QoS demand \cite{ jain2021towards}. However, single UAV systems cannot meet the requirements of advanced surveillance due to limited resources and computational power, as surveillance drones are required to stay in the air for hours or days, and their high-tech cameras need to scan the entire city and zoom in for advanced monitoring. Moreover, UAVs also require the integration of complex machine-learning algorithms for video analytical operations to extract useful information for independent decision-making \cite{ ishtiaq2021edge }. Nevertheless, single UAV systems' limited resources and computational power significantly hinder their potential for advanced surveillance.

At the same time, multiple UAVs equipped with different types of high-resolution cameras, live-feed cameras, infrared cameras, radar, and sensing components can be assigned diverse tasks to provide advanced collaborative surveillance for various commercial, civil and military applications. For example, military drones with integrated advanced live-feed cameras and sensors can enable continuous surveillance for longer hours. In contrast, UAVs with more storage and computational power can provide edge services to process the monitored information with high QoS and low resource consumption. In addition, multiple UAVs can work collaboratively as a high-level perimeter and a response system to prevent unauthorized access in industrial plants such as factories, solar parks, offshore platforms, quarries, and refineries. The images collected from multi-UAV systems are vital for reconnaissance or rapid situation awareness. Ground security can also use ariel surveillance to detect and monitor potential threats from a safe distance and reduce the requirement for foot patrols by security guards. Maritime defense can also significantly benefit from the multiple-UAVs system for counter-piracy operations to monitor, analyze and anticipate pirate vessel movements for identifying hazardous locations and evaluating piracy trends around the world. Moreover, UAVs can also work collaboratively to provide protection services for superyachts, such as helping crews stay safe and reducing delays and cost overruns, which minimizes insurance premiums while assuring stakeholders that company assets are secure.

\subsection{Cooperative Aerial Imaging for Remote Sensing}
Numerous military, commercial and civilian applications require aerial imaging for various purposes such as homeland security, border patrol, monitoring forest fires, tracking wildlife, and nuclear power plants perimeter monitoring \cite{beard2006decentralized}. UAVs integrated with high-resolution cameras can perform cooperative aerial imaging by capturing individual images from different viewpoints. Later, the captured images can be analyzed separately or combined to create an overall image. In addition, pictures taken from multiple UAVs flying at a low altitude provide vital information compared to the pictures captured from helicopters or airplanes. UAVs with integrated multimodal sensors and improved photogrammetry and computer algorithms have also displayed great results for acquiring and processing terrain data. The digital surface models and digital elevation models generated from drones provide essential inputs for topography for the accurate modeling of flood plain hydrodynamics, and overland flow predictions \cite{grau2021improved}.
Moreover, UAVs equipped with multispectral cameras also provide the most accurate metadata, leading to efficient and straightforward imagery data collection for vegetation mapping \cite{ezequiel2014uav}. Accordingly, a swarm of UAVs can cooperatively cover a large area to provide a synoptic view efficiently and economically. UAVs can also provide data under clouds, which is particularly useful in tropical areas where cloud cover is frequent for long periods of the year. Thus, UAVs can improve satellite data's spatial, temporal or spectral resolutions by providing a complementary dimension through fusion methods.
\begin{figure}
\center
\includegraphics[width=3.0in, height = 6.0cm]{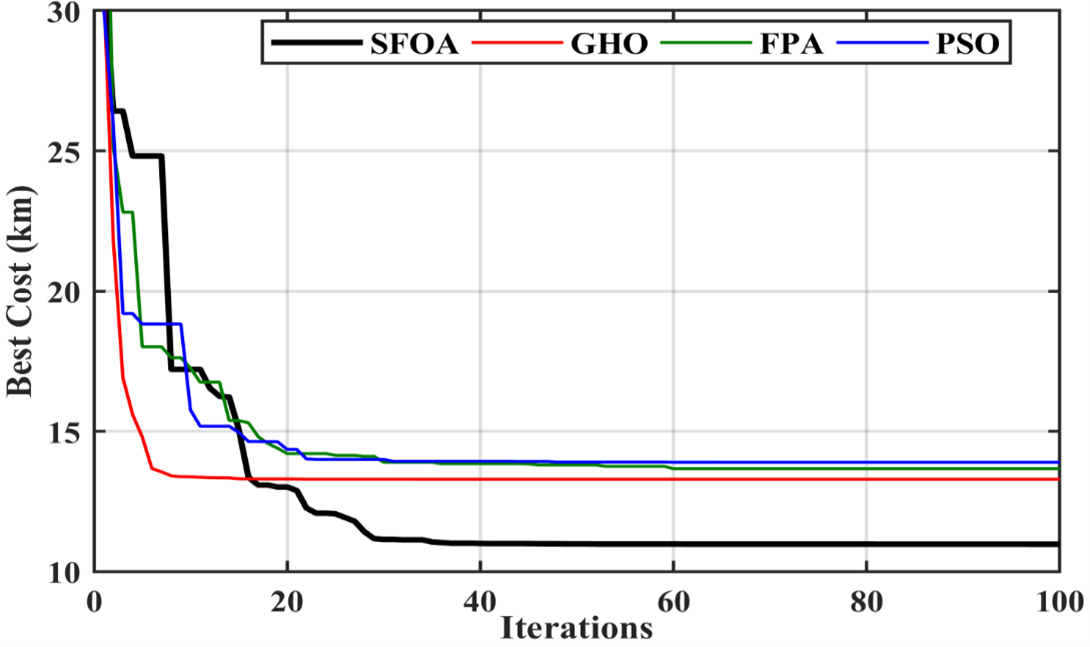}
\vspace{2.5 mm}
\caption{Cost vs. number of iterations for UAVs-based disaster management.}
\label{Cost}
\vspace{-4 mm}
\end{figure}
\begin{figure*}[t]
\center
\includegraphics[width=5in, height = 7cm]{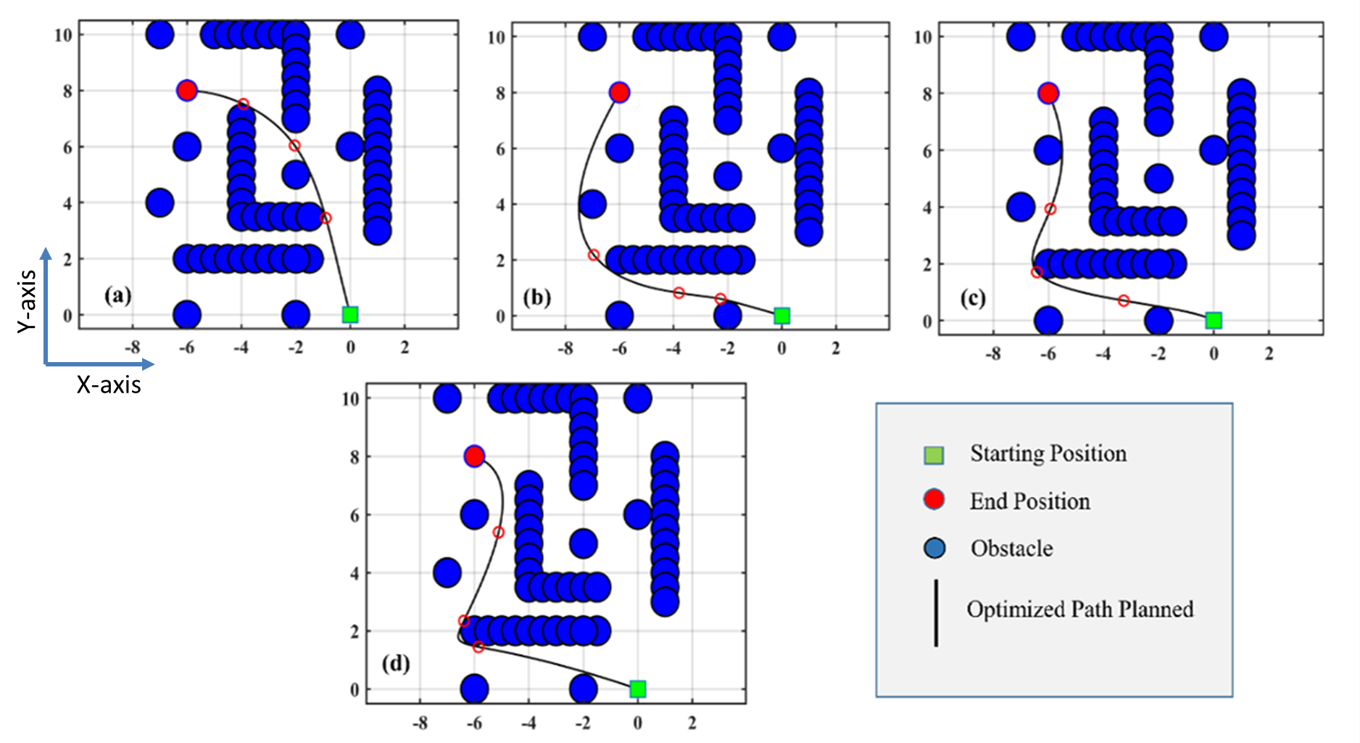}
\vspace{2.5 mm}
\caption{Comparison between metaheuristic algorithms (a)SFOA (b)GHO (c)FPA (d)PSO.}
\label{Comp}
\vspace{-4 mm}
\end{figure*}

\subsection{Disaster Management}
UAVs have shown significant strides and advances in disaster risk reduction, preparedness, response, recovery, relief, and rehabilitation. A swarm of UAVs can provide better help and support during a disaster by assigning different recuse and relief tasks to a group of UAVs. Accordingly, sufficient knowledge and resource sharing among UAVs can better facilitate the victims with medical aid and rescue operations. For example, drone-led surveillance can detect a potential or apparent disaster quickly by flying to a potential hazard to investigate a suspicious occurrence. Collaborative UAV-based reconnaissance gives a first-hand analysis of the disaster point and helps strategize on-ground action much faster than manual detection, investigation, and action. In addition to emergency response and disaster awareness, a group of drones can efficiently perform supply-drop runs with medical and essential packages during floods, cyclones, or earthquakes for the well-being of stranded survivors. For example, UAVs can be notified by the other surveillance providing UAVs to deliver medical equipment such as Automated External Defibrillators (AED) (i.e., AED help analyze the heart rhythm and deliver an electric shock to victims of ventricular fibrillation to restore the heart rhythm to normal) for first aid.

Furthermore, the collaborative working of UAVs also exhibits a tangible impact on disaster response by performing multiple search and rescue operations. Drones equipped with high-definition cameras and thermal sensors can conduct high-resolution visual and thermal imaging to clearly picture survivors, even under rubble or within inaccessible crevices. The information provided by the drones helps the on-ground team to understand all access points and paths toward the survivors, assisting in better understanding the current situation in terms of communication or transport disruption. Recently, Qadir et al. \cite{qadir2022optimizing} compared four state-of-the-art metaheuristic algorithms for disaster management using UAV swarm in the context of smart cities. The algorithms are designed in such a way to provide obstacle-free efficient paths for UAVs in a complex environment. Figure \ref{Cost} illustrates the best cost vs. iteration comparison between Particle Swarm Optimization (PSO), Flower Pollination Algorithm (FPA), Grasshopper Optimization (GHO), and Smart Flower Optimization Algorithm (SFOA). Figure \ref{Comp} shows the trajectory optimization between these algorithms from start to endpoints having obstacles in the complex environment. For this complex case, Table \ref{ComTable} shows that the SFOA algorithm outperforms the other algorithms having a computation time of 109.42 s with the distance covered as 10.971 km.

%

%

%


\begin{table}[!]
\centering
\caption{Comparison for the complex case }
\label{ComTable}
\resizebox{9cm}{!}{%
\begin{tabular}{@{}ccc@{}}
\toprule
\multirow{2}{*}{Metaheuristic algorithm} & \multicolumn{2}{c}{Complex Case}          \\ \cmidrule(l){2-3}
                                         & Path length (km) & Computational t (s) \\ \midrule
PSO                                      & 14.547           & 126.210                \\ \midrule
FPA                                      & 13.892           & 119.191                \\ \midrule
GHO                                      & 13.2889          & 113.568                \\ \midrule
SFOA                                     & 10.9719          & 109.42                 \\ \bottomrule
\end{tabular}%
}
\end{table}

\subsection{Connecting the Unconnected}
In the digital world, every individual should be able to live, learn, work, and participate. However, there is a gap between regions with access to communication services and those without restricted access. Hence, the digital divide must be bridged to make the world more resilient and inclusive. The situation is much worse in the least developed countries, where an average of two out of every ten people have connectivity. The question is how to reach people at risk of being left behind? Recent research in telecommunication recommends the development of aerial networks in rural areas to connect the unconnected since connectivity is no longer a luxury but a lifeline. In this context, the swarm of UAVs is a viable solution to provide coverage in unconnected regions, creating a flying ad hoc network (FANET) \cite{s21238037}. Bringing aerial networks-based connectivity, especially in rural and underserved areas, can enable various applications, such as online education, smart agriculture, e-health, and an efficient supply chain.

In addition to the wide range of opportunities in rural areas, FANET also brings significant benefits in realizing the smart city concept by enabling seamless connectivity. UAV-assisted applications for intelligent traffic monitoring, key infrastructures management and monitoring development work regularly can enhance the lives of its residents by providing efficient infrastructure and services at a reduced cost \cite{menouar2017uav, hernandez2018internet, mohamed2017uavfog}. For example, UAVs can considerably influence the construction industry by incorporating new X-ray technology to help craft high-resolution 3D maps of objects through Wi-Fi. X-ray technology can also create models of areas hidden behind walls or other barriers to give workers more important information to monitor dangerous structures that could harm workers or the general public. Moreover, collaborative UAVs can help in the recovery of the network in case of disasters \cite{Matracia2022} and can also improve coverage for underserved regions, such as maritime networks \cite{Alqurashi2022}.
\begin{table*} [!]
\renewcommand{\arraystretch}{}
\noindent\begin{minipage}{\linewidth}
\centering
\captionof{table}{{Summary of future research directions.}}
\begin{tabular}{ p{3cm} p{4cm} p{4cm} p{4cm} }
\label{Table 3}
\\\hline
Author  Year                    & Contribution  & Challenges  & Possible Future Direction
\\ \hline
Qadir et al., 2022 \cite{qadir2022optimizing}            & Trajectory Optimization and Scheduling                                        & Computation time and complexity        & Proposing different metaheuristic algorithms for UAV trajectory optimization                                      \\
Nguyen et al., 2021 \cite{9415623}            & Integration of Federated learning                                        & Performance accuracy and computational complexity for machine learning algorithms         & Incorporating resource aware Federated Learning architecture for minimizing computational complexity
\\
Wang et al., 2022 \cite{wang2022deployment}            & Cellular Integration                                        &  Deployment of UAV and association schemes        & Enhancing the battery life by minimizing the transmission power consumption of UAVs
\\
Tedeschi et al., 2022 \cite{tedeschi2022ppca}            & Self-organization                                        & Collisions with obstacles, neighbouring UAVs and data privacy       & Privacy preserving scheme along with collision avoidance algorithms for secure and reliable information sharing
\\
Sahni et al., 2021 \cite{9223724}            & Collaborative task completion (Collaborative Task Offloading for edge computing)                                        & Network congestion and performance delay        & Implementing the collaborative edge computing (CEC) for computational resource optimization between different edge devices and sharing data securely
\\

Yang et al., 2021 \cite{9264742}            & Energy-efficiency                                        & Wireless transmission and local computation        & Developing Hybrid iterative algorithm along with the proactive approach for power optimization aiding wireless transmission
%\\
%Shi et al., 2021 \cite{shi2021review}            & Middleware requirement for UAV collaboration                                        & Collaboration, task and information sharing between UAVs         & Enhancing the performance, structure and communication middleware for robust UAV collaboration
%\\
%Qadir et al., 2021 \cite{qadir2021addressing}            & UAV-to-UAV/ UAV-to-Ground Link Design                                        & Data rate and network coverage        & Designing B5G/6G network will provide higher data rate along with vast coverage area for several applications
\\

\\ \hline
\end{tabular}
\end{minipage}
\end{table*}
\section{Future Research Directions}  \label{directions}
Collaborative UAVs are the new future and can play a significant role in serving different real-world applications. In this article, we reviewed various aspects of collaborative UAVs, including communication, control, and collaboration requirements. However, there are still many issues with establishing an efficient swarm of UAVs to perform coordinated tasking. Therefore, in this section, we highlight a few exciting future directions as summarized in Table \ref{Table 3}.

\subsection{Trajectory Optimization and Scheduling}
UAV trajectory optimization in a complex environment depends on the cost associated with the flight and the time it takes to reach the destination. In the dense urban environment, UAV trajectory planning needs to consider several important issues (e.g., signal multipath interference, obstruction, or attenuation) as UAVs fly at relatively lower altitude in the presence of many natural and man-made obstacles \cite{causa2021multiple}. Moreover, trajectory planning in complex environments also faces various challenges from the navigation point of view that compromises mission autonomy, control and planning functionalities that degrade UAV performance for smart city applications such as environmental inspection and monitoring, mobility of people and goods and traffic management. Therefore, the challenges associated with trajectory optimization and scheduling (such as accurate prediction of UAV trajectory, modeling the safety of UAV flight, and the ordering and timing decisions for UAV) need to be mathematically modeled to minimize the traveling time and battery consumption. In this regard, metaheuristic algorithms can be investigated for UAV trajectory optimization and time scheduling for future insights.

\subsection{Integration of Federated learning}
Machine learning (ML) is an emerging field contributing to several applications in collaborative UAV tasks. ML, a sub-field of Artificial Intelligence (AI), uses state-of-the-art techniques to design algorithms that work closely with human nature and are more efficient.
Owing to the diversified applications of ML, the designed algorithms are time-consuming and computationally complex as well. To address these constraints, there is a need to develop resource-aware Federated Learning (FL) architecture that can address computational complexity, limited onboard memory, and high latency limitations for collaborative UAVs.
Moreover, collaborative UAVs will provide seamless and reliable communication while distributing the tasks, requiring an energy-efficient approach. In this context, trajectory optimization and the role of FL need to be further researched in helping to create a practical framework where UAVs can collaborate without any energy constraint, which is still a big challenge open for further investigation.

\subsection{Cellular Integration}
UAV deployment and alternative association schemes are the real backbones for providing UAV-integrated cellular coverage. In the dense urban environment where network connectivity is affected by tall buildings, bridges and tunnels, UAV-integrated cellular coverage can enhance the network connectivity for Vehicular ad hoc networks (VANETs) and optimize the selection of the fastest route. While during a natural disaster, when the communication architecture is destroyed, or no other modes of communication are available, UAV-integrated cellular coverage can provide a self-sustaining infrastructure to provide surveillance or relief services. However, for large-scale cellular integration, it is necessary to understand the bandwidth, range, speed, cost, and power requirements of the communication technology to enable UAV-integrated cellular networks \cite{sharma2020communication}. Existing studies also emphasized investigating low latency and high data rate ultra-reliable communication requirements for collaborative UAV communication.

\subsection{Self-organization}
Self-organization and automation in collaborative UAVs can perform data aggregation, management, privacy preservation, and path optimization. In complex urban environments, the use of UAV-based applications is exponentially growing for every domain of life as UAVs with self-organizing capabilities can develop a self-sustainable network to complete complex operations in reduced time with better efficiency. For example, a swarm of UAVs can self-organize to work independently to monitor a large agricultural field without taking frequent instructions from the central base station. Similarly, UAVs with self-organizing capabilities can provide cellular coverage or disaster relief services to help the victims during a natural disaster. However, achieving self-organization in a swarm of UAVs using distributed, hybrid and centralized methods is quite challenging. For instance, in a confined environment where UAVs have to hover and collect data for processing and analyzing to make collaborative decisions is a demanding task and needs further investigation. Furthermore, collision avoidance among neighboring UAVs, obstacle avoidance, and data are also daunting challenges that must be addressed. Towards this end, state-of-the-art collision avoidance algorithms integrated with privacy preservation schemes must be adopted for self-organization in the swarm of UAVs.

\subsection{Collaborative Task Offloading}
Task offloading is a serious concern for many applications dealing with different activities in a dedicated space or a complex environment. Nevertheless, it results in performance delays and network congestion. However, collaborative edge computing (CEC) can reduce computational costs and securely transfer data between edge devices \cite{ishtiaq2021edge}. For example, in the digital healthcare environment, edge-computing services help in analyzing the patient's data at the local server with low latency and security concerns. Accordingly, UAV-based task offloading allows UAVs to provide edge computing services for improved computational efficiency, QoS and reduced latency. Task offloading using UAVs, especially in complex environments, suffers from signal coverage problems that can be solved by optimizing the UAVs' mobility and improved collaboration strategies. Also, the UAVs can collaborate for task offloading using edge computing or transmitting it to the ground station to further improve the model efficiency.

\subsection{Energy-efficiency }
Energy scarcity is one of the major concerns for next-generation wireless communication systems. In a collaborative UAV network, the UAVs transmit data and retrieve information at the energy cost consumed by batteries. For example, UAV-based long-term surveillance requires UAVs placement for longer hours, leading to a massive amount of video data generation that needs to be transmitted, consuming considerable power resources. Thanks to the emerging concept of Tethered UAVs, the battery problem is now solved at the expense of limited mobility and coverage. However, in various UAV-based applications for emergencies, energy is still a primary issue; for instance, during disaster management, UAVs are required to monitor, scan, and collect videos/images to notify the authorities for relief operations. Thus, they need enough power resources to enable efficient management and rescue operations. Moreover, developing optimization algorithms that optimize the hovering, sensing time, and power for UAVs is needed to further improve the network's energy efficiency.

%\subsection{Middleware requirement for UAV collaboration}
%Middleware is like a middleman that can be a driver, translator, operating system depending on the application where it is needed. Collaborative UAVs owns unique challenges highlighted by their heterogeneity, reliability, high mobility, stability and flying related challenges. For this purpose, middleware can play a significant role in addressing these challenges.

%Based on previous research, Middleware can solely offer several advantages in developing UAV systems. However, in the perspective of future directions, it must add more functionality and common architecture to be used in collaborative environment without the need to change UAV applications.

%\subsection{UAV Connectivity}
%UAV to UAV and UAV to Ground linkage has a vast potential in civilian and military applications associated with higher data rate and network coverage. The flexibility of these UAVs make them more popular in agriculture, meteorological, medical, and many other applications. Referring to these application, UAV-UAV, UAV-Terrestrial, UAV-Satellite and UAV-HAP communication can provide afficacy in terms of coverage and data transmission.

%In case of any disaster management, the telecommunication infrastructure can be destroyed, therefore mobile station aided with B5G/6G network will provide a vast coverage area along with higher data rate for collaborative UAVs.

\section{Conclusion} \label{conclusion}
The importance of collaborative UAVs has proliferated to advance their autonomy and coordination for a wide range of applications. At the same time, developing collaborative UAVs faces various challenges concerning communication, control, and collective decision-making. Therefore, this review thoroughly studies the potential and challenges of collaborative UAVs to highlight the state-of-the-art developments. We comprehensively discuss the existing literature to summarize collaborative tasks such as trajectory formation, target localization, data collection, and cooperative decisions that advance the multi-UAV system's performance. Moreover, we also explore the real-world application of UAV systems and highlight their role in advanced monitoring, surveillance, and management. Towards the end, we provide possible future directions that need attention for realizing the broader concept of collaborative UAVs. In a nutshell, this review offers an in-depth discussion for researchers and engineers to understand the collaborative aspect of a swarm of UAVs for designing an efficient network architecture.
\section{Acknowledgments}
This research is supported by the National Key Research and Development Program of China under Grant 2020AAA0108905; in part by the National Natural Science Foundation of China under Grant 61825303, Grant 62088101; in part by the Shanghai Municipal Commission of Science and Technology Project under Grant 19511132101; in part by the Shanghai Municipal Science and Technology Major Project under Grant 2021SHZDZX0100.

\bibliographystyle{IEEEtran}
\bibliography{ref.tex}

\end{document}